\documentclass[final]{aipproc}

\layoutstyle{8x11single}

\SetInternalRegister\hbadness{8000} 

\newcommand\doingARLO[2][]{%
  \ifx\mmref\undefined #1\else #2\fi
}

\def\be {\begin{equation}}
\def\ee {\end{equation}}
\def\bea {\begin{eqnarray}}
\def\eea {\end{eqnarray}}

\def\2l {\frac{{f_i}}{(2\lambda + 1)}}

\begin{document}

\title{Squeezing lepton pairs out of broken symmetries}


\author{A. K. Dutt-Mazumder}{
  address={Department of Physics, McGill University, Montreal, QC, 
Canada H3A 2T8},
  email={abhee@physics.mcgill.ca},
}

\author{C. Gale}{
  address={Department of Physics, McGill University, Montreal, QC, 
Canada H3A 2T8},
  email={gale@physics.mcgill.ca},
}

\author{A. Majumder}{
  address={Department of Physics, McGill University, Montreal, QC, 
Canada H3A 2T8},
  email={majumder@physics.mcgill.ca},
}

\author{O. Teodorescu}{
  address={Department of Physics, McGill University, Montreal, QC, 
Canada H3A 2T8},
  email={octavian@physics.mcgill.ca},
}

\begin{abstract}
We discuss two possible signatures of symmetry breaking that can 
appear in dilepton spectra, as measured in relativistic heavy ion
collisions. The first involves scalar-vector meson mixing and is 
related to the
breaking of Lorentz symmetry by a hot medium. The second is related to
the breaking of Furry's theorem by a charged quark-gluon
plasma. Those signals will be accessible to upcoming measurements to be
performed at the GSI,  RHIC, and the LHC.  

\end{abstract}


\maketitle
\vspace*{-1cm}
\section{Introduction}

Electromagnetic radiation, especially lepton pairs, represents probes
of choice for the study of heavy ion reactions. Owing to the
smallness of their final state interaction they travel essentially
unscathed from creation to detection. Furthermore, via Vector
Meson Dominance (VMD) \cite{sak}, real and virtual photons are ideal 
as signatures of the in-medium behaviour of vector mesons \cite{gal91}. 
This aspect has been highlighted in a theoretical interpretation
\cite{rawa} of the
low dilepton invariant mass measurements performed by the CERES
collaboration \cite{ceres}. In short, it appears that the behaviour of
hot and dense strongly interacting matter can indeed be revealed by 
electromagnetic observables \cite{gaqm}. 

In this work we shall outline how symmetry breaking effects can
manifest themselves in dilepton spectra. The first of those involve the
possibility of $\rho - a_0$ mixing in dense nuclear matter. The size of
this contribution is density-dependent and goes to zero in vacuum, on
account of Lorentz symmetry. We show that such a mixing opens up a new  
channel in dilepton production and will modify the spectrum in the
$\phi$ invariant mass region. Secondly, we show that Furry's theorem is
no longer obeyed in a baryon rich quark gluon plasma. This has the
consequence of permitting new processes like $g g \to l^+ l^-$. We
evaluate this contribution and compare it with $q \bar{q} \to l^+ l^-$.
We will consider only dielectrons here and the details of the
calculations will only be sketched; the interested reader is invited to
consult the references quoted in the text. 
The observables we discuss in this work are within reach of
measurements soon to be performed by the HADES collaboration at the
GSI, and by the PHENIX collaboration at RHIC. 

\vspace*{-0.5cm}
\section{Dileptons from meson mixing in dense hadronic matter}

There exists physical processes which
are forbidden in free space but can take place in matter.
Those are related to medium-induced symmetry breaking effects.
The interaction respects the symmetry, but it is broken by the ground
state.
For instance, in matter, Lorentz symmetry is lost which leads to the mixing
of different spin states even when the interaction Lagrangian respects
all the required symmetry properties.
A well-known example
of this is $\sigma$-$\omega$ mixing in nuclear matter
as discussed by Chin \cite{chin77}. 
Another manifestation lies in the fact that different polarization
states of vector fields can have different dispersion relations
\cite{gal91}. 

Our goal here is to identify the effects of scalar
and vector meson mixing on dilepton production rates,  which could 
be observed in high energy heavy ion experiments.
The  mixing of mesons considered here include both isoscalar
and isovector channels.
To be more specific, we estimate the rate of dilepton production from
$\sigma$-$\omega$ and $\rho$-$a_0$ mixing. Some attention
was paid to the former in the context of heavy-ion collisions
\cite{wolf98,saito98}, and the importance of the later was shown
recently \cite{teodorescu00}.
 
In this first part we present thermal
production rates of dileptons induced by scalar-vector mixing in the
isoscalar and isovector channels.
Those results are 
compared against $\pi$-$\pi$, $K$-${\bar K}$
annihilation contributions, which are treated here as standard candles. 
Note that processes involving mixing of different G-parity states
like $\sigma$-$\omega$ or $\rho$-$a_0$
are allowed only in matter which is not
invariant under charge conjugation. Therefore,
$\pi \pi$ $\rightarrow$ $e^+e^-$ through the coupling to nucleons
via $\sigma$ and $\omega$, or $\pi \eta \rightarrow e^+e^-$ mediated
by $\rho$-$a_0$ mixing,
can take place in matter
with a finite nucleon chemical potential. Hence it is natural to look
for such density-driven effects in experiments involving
relatively low energy collisions where one expects to have a 
relatively large chemical potential.

The interaction Lagrangian used in the present model can be written as
\bea
{\cal L}_{int} = g_\sigma {\bar \psi}\phi_\sigma \psi +
          g_{a_0} {\bar \psi}\phi_{a_0,\alpha}\tau^\alpha \psi
          + g_{\omega NN}{\bar{\psi}} \gamma _\mu\psi\omega^\mu
          + \\
 g_{\rho} [{\bar{\psi}} \gamma _\mu \tau^\alpha
 \psi + \frac{\kappa _\rho}{2m_n}{\bar{\psi}}
    \sigma_{\mu\nu}\tau^\alpha \partial ^\nu] \rho^\mu_\alpha\ ,
\eea
where $\psi$, $\phi_\sigma$, $\phi_{a_0}$, $\rho$ and $\omega$ correspond
to nucleon, $\sigma$, $a_0$ , $\rho$ and $\omega$ fields, and the $\tau$'s are
Pauli matrices. From this point on we use s and v to denote scalar
and vector mesons, {\em i.e.} s = $\sigma$, $a_0$ and v = $\omega$, $\rho$. It 
is understood that $\kappa_\omega=0.0$.

The polarization vector through which the scalar meson couples to the
vector meson via the n-n loop is given by

\bea
\Pi_ \mu (q_0,|{\vec q}|) &=& 2i g_s g_v \int \frac{d^4k}{(2\pi)^4}
    \mbox{Tr}[G(k) \Gamma_\mu G(k+q)]. \label{pim}
\eea
where 2 is an isospin factor and the vertex for v-nn
coupling is:
\bea
\Gamma_\mu=\gamma_\mu - \frac{\kappa_v}{2m_n}\sigma_{\mu\nu}q^\nu\ .
\label{vertex}
\eea
$G(k)$ is the in-medium nucleon propagator \cite{sewal}. 

\begin{figure}[htbp]
  \resizebox{18pc}{!}{\includegraphics[height=0.5\textheight]{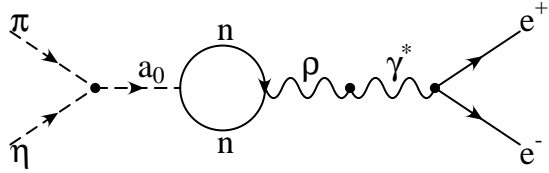}}
 \caption
  {\small The Feynman  diagram for the process $\pi + \eta
\rightarrow e^+ + e^-$. \label{fig:loop}}
\end{figure}

$\rho$-$a_0$ mixing opens up
a new channel for dilepton production in dense and hot nuclear matter:
$\pi + \eta \rightarrow e^+ + e^-$, accessible  
through n-n excitations. The Feynman diagram for this 
process is  shown in Fig.~\ref{fig:loop}. Similarly, the contribution
for $\pi + \pi \rightarrow \sigma \rightarrow \omega \rightarrow e^+ + e^-$ 
can be evaluated from a diagram obtained from that in
Fig.~\ref{fig:loop} by replacing one of the incoming $\pi$'s by $\eta$
and by performing the following substitutions: $a_0 \to \sigma$ and
$\rho \to \omega$.

\vspace*{-0.5cm}
\subsection{Dilepton rates}

Once we obtain the relevant cross-section for the dilepton production 
channel, the
thermal production rate of the lepton pairs
(number of reactions per unit time, per unit volume
$R_{12}^{e^+e^-}=dN_{e^+e^- pairs}/d^4x$) can be estimated
in the independent particle approximation of relativistic kinetic theory:

\bea
\frac{dR_{12}^{e^+e^-}}{dM^2}=\int \frac{d^3k_1}{(2\pi)^3}f_1({\bf k}_1)
\int \frac{d^3k_2}{(2\pi)^3}f_2({\bf k}_2)
\frac{d\sigma_{12}^{e^+e^-}}{dM^2}(s,M^2)v_{rel}
\label{rate1}
\eea
where $f_1,f_2$ are the thermal distributions of the 1,2~species and
$v_{rel}=\frac{\lambda^{1/2}(s,m_1^2,m_2^2) \sqrt{s}}{2 E_1 E_2}$ with
$\lambda(x,y,z)=x^2+y^2+z^2-2xy-2xz-2yz$ is the triangle function.
%
%
%
%

\begin{figure}[htbp]
  \resizebox{18pc}{!}{\includegraphics[height=0.8\textheight]{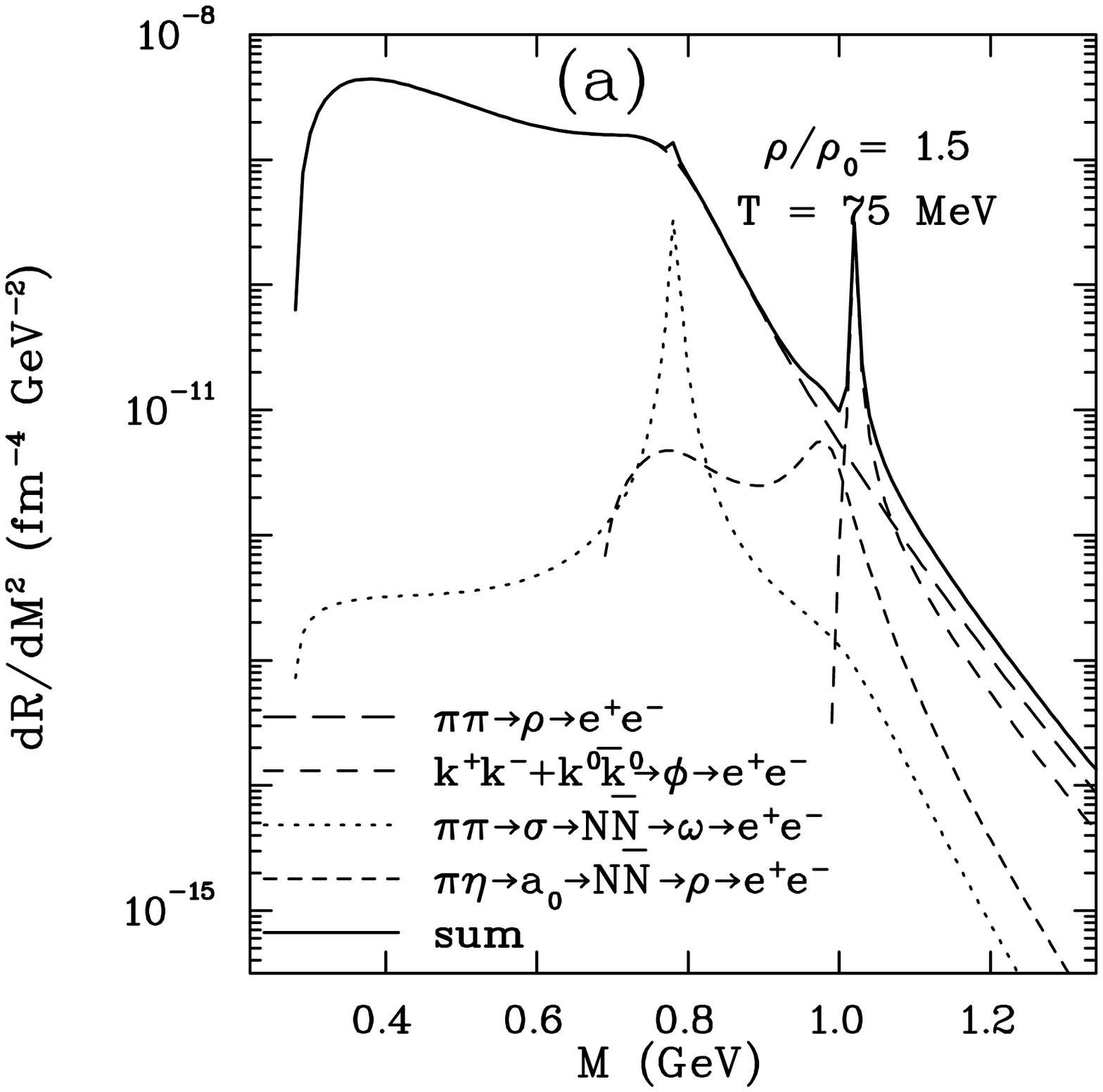}}
  \resizebox{18pc}{!}{\includegraphics[height=0.8\textheight]{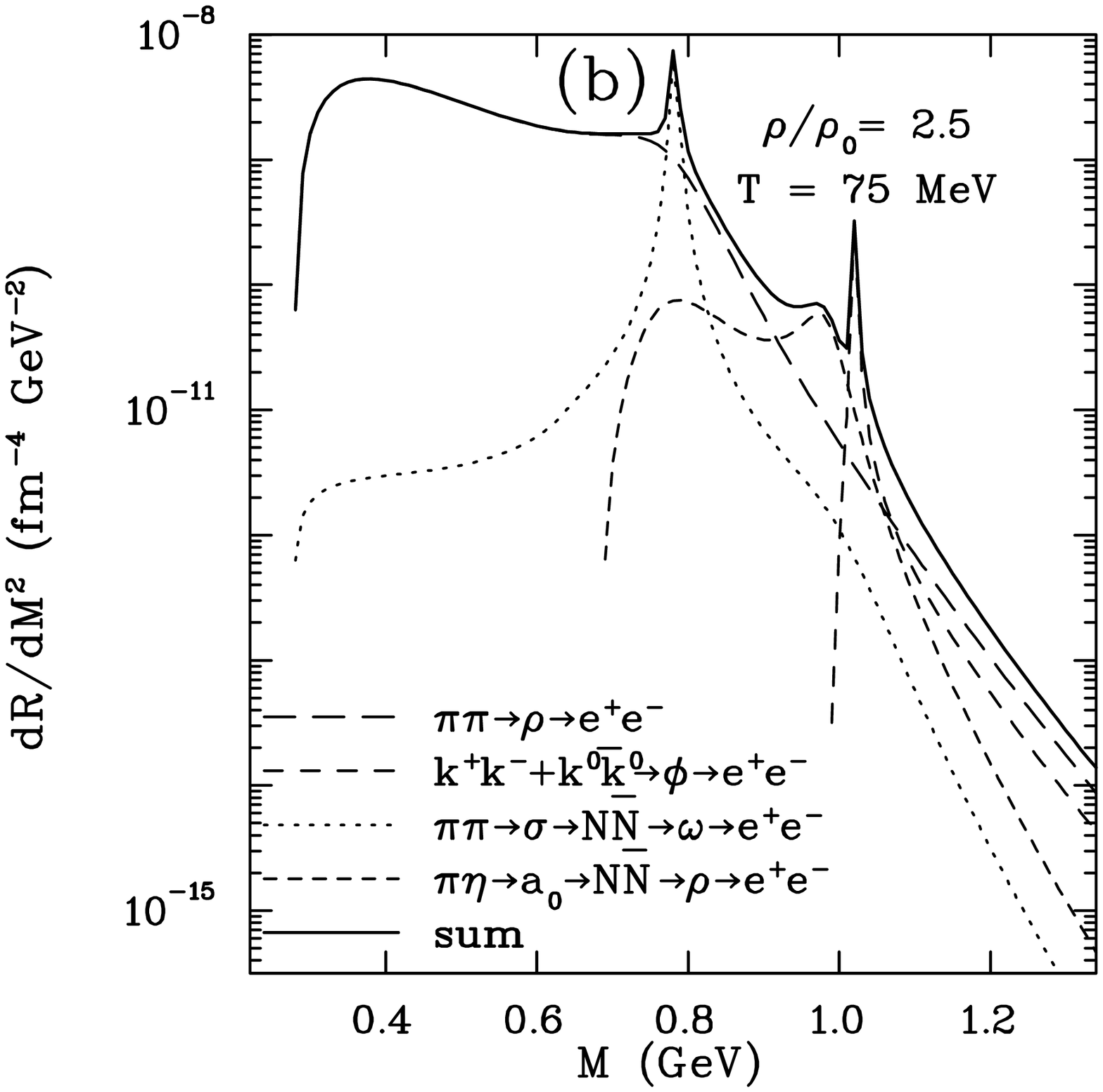}}
 \caption
  {\small Rates at T = 75 MeV with mixing in both isoscalar and isovector
channels.
\label{fig:spectra75}}
\end{figure}

Fig.~\ref{fig:spectra75} shows the  total dilepton yield with the combined 
effect of mixing in the isoscalar and isovector channels. We have
chosen conditions that are representative of collisions to occur at
the GSI, and we show the pion and kaon annihilation channels
separately.  It is 
important to see that rates induced by the mixing can be significant
in certain windows of density and temperature.
Particularly, mixing induces a higher yield of dileptons in the invariant mass 
region
between the $\rho$ and $\phi$ peaks, and is strongly density dependent. 
The mixing in the isoscalar
channel contribute essentially at the $\omega$ peak
\footnote{Technically, the $\pi \pi$ annihilation channel should
treat the $s-$ and the $p-$wave annihilation coherently. The interference
effects have been found to be negligible.}. 
On the other hand, the isovector channel seems to provide a
better probe as it contributes in the tail region of the two pion
annihilation contribution. It should be stressed here that broadening or
dropping $\rho$ meson mass would even favour our observation. In that
case the $\pi$-$\pi$ background would be pushed more towards lower
invariant mass region bringing the $a_0$ peak into a clearer relief.
Importantly, the upcoming HADES experiment \cite{hades} has the necessary
resolution, and will operate in the appropriate energy regime to
adequately investigate the physics discussed here.


\vspace*{-0.5cm}
\section{Dilepton emission from a baryon-rich quark gluon plasma}

At zero
temperature, 
diagrams in QED that contain a fermion loop with an odd number of
photon vertices are cancelled by an equal and opposite 
contribution coming from the same diagram with fermion lines running in
the opposite direction (Furry's theorem \cite{fur37}). This statement can 
also be generalized to QCD for processes with two gluons and an odd 
number of photon vertices.

%
%
 
At finite temperature and density, the corresponding quantity to consider in
connection with Furry's theorem is 
$\sum_{n} \langle n| A_{\mu_1} A_{\mu_2} ... A_{\mu_{2n+1}} |n\rangle 
e^{-\beta (E_n - \mu Q_n)},
$
where $\beta$ = 1/T, $\mu$ is a chemical potential, and $A_\mu$ is an
electromagnetic gauge field operator. 
Here, the effect of the charge conjugation operator $C$ is such that  
$C|n\rangle = e^{i\phi}|-n\rangle$, where $|-n\rangle$  is a state in 
the ensemble with
the same number of antiparticles as there are particles in $|n\rangle$ 
and vice-versa.
If $\mu = 0$,  
upon multiple insertions of the unit operator $C^{-1}C$ we get

\vspace{-0.5cm} 

\begin{eqnarray}
\langle n| A_{\mu_1} A_{\mu_2} ... A_{\mu_{2n+1}} |n\rangle 
e^{-\beta E_n}
&=& - \langle -n| A_{\mu_1} A_{\mu_2} ... A_{\mu_{2n+1}} |-n\rangle 
e^{-\beta E_n}. 
\end{eqnarray}

\noindent The sum over all states will contain the mirror term 
$\langle -n| A_{\mu_1} A_{\mu_2} ... A_{\mu_{2n+1}} 
|-n\rangle e^{-\beta E_n} $ with
the same thermal weight. Therefore 
%
%
%
\begin{eqnarray}
\sum_{n} \langle n| A_{\mu_1} A_{\mu_2} ... A_{\mu_{2n+1}} |n\rangle 
e^{-\beta E_n } = 0\ 
\end{eqnarray}
and Furry's theorem still holds. 

However, if
$\mu \neq 0$ 
(unequal number of particles and antiparticles), 
then
%
%
\begin{eqnarray}
\langle n| A_{\mu_1} A_{\mu_2} ... A_{\mu_{2n+1}} |n\rangle 
e^{-\beta (E_n - \mu Q_n)}
= - \langle -n| A_{\mu_1} A_{\mu_2} ... A_{\mu_{2n+1}} |-n\rangle 
e^{-\beta (E_n - \mu Q_n)}\ .
\end{eqnarray}
The mirror term this time is 
$
 \langle -n| A_{\mu_1} A_{\mu_2} ... A_{\mu_{2n+1}} |-n\rangle 
e^{-\beta (E_n + \mu Q_n)}$, with a different thermal weight. Thus 
%
%
\begin{eqnarray}
\sum_{n} \langle n| A_{\mu_1} A_{\mu_2} ... A_{\mu_{2n+1}} |n\rangle 
e^{-\beta (E_n - \mu Q_n)} \neq 0\ .
\end{eqnarray}
One may say that the medium, being charged, manifestly breaks charge 
conjugation invariance and these
 Green's functions are thus finite and will lead to the appearance of new processes in
a perturbative expansion. Again, the appearance of processes that 
can be related to symmetry breaking in a medium 
  has been noted before \cite{chin77,wel92}.


\begin{figure}[htbp]
  \resizebox{18pc}{!}{\includegraphics[height=.2\textheight]{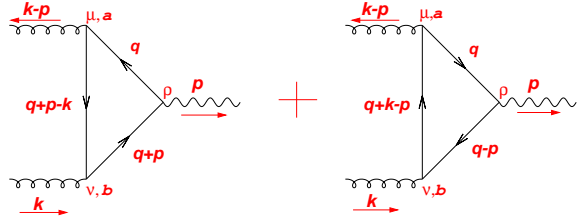}}
    \caption{ The two gluon photon effective vertex as the sum of two diagrams
with quark number running in opposite directions. }
    \label{2vert}
\end{figure}

Let us now focus our attention on the diagrams of Fig. \ref{2vert} 
for the case of two gluons and a photon attached to a quark loop. 
Such a 
process does not exist at zero temperature or even at finite 
temperature and zero 
density. At finite density this leads to a new source of dileptons: $g
g \to e^+ e^-$.
In order to obtain the full matrix element one 
must coherently sum contributions from both diagrams which have fermion number
running in opposite directions. The amplitudes for 
${\mathcal T}^{\mu \rho \nu} (=  T^{\mu \rho \nu} +  T^{\nu \rho \mu})$  are

\vspace{-0.5cm} 

\begin{eqnarray}
\!\!\!\!\!T^{\mu \rho \nu} &=& \frac{1}{\beta} \sum_{n= -\infty}^{\infty}
\int^{\infty}_{-\infty} eg^{2} tr[t^{a}t^{b}]
\frac{d^{3}q}{(2\pi)^{3}} Tr[ \gamma^{\mu} \gamma^{\beta} \gamma^{\rho} 
\gamma^{\delta} \gamma^{\nu} \gamma^{\alpha}]  
\frac{(q+p-k)_{\alpha} q_{\beta} 
(q+p)_{\delta} }{(q+p-k)^{2} q^{2} (q+p)^{2}}, \nonumber
\end{eqnarray}    

\begin{eqnarray}
\!\!\!\!\!T^{\nu \rho \mu} &=& \frac{1}{\beta} \sum_{n= -\infty}^{\infty}
 \int^{\infty}_{-\infty} eg^{2} tr[t^{a}t^{b}]
\frac{d^{3}q}{(2\pi)^{3}} Tr[ \gamma^{\nu} \gamma^{\delta} \gamma^{\rho} 
\gamma^{\beta} \gamma^{\mu} \gamma^{\alpha}] 
\frac{(q+k-p)_{\alpha} q_{\beta} 
(q-p)_{\delta} }{(q+k-p)^{2} q^{2} (q-p)^{2}}\ , \label{2g1p}
\end{eqnarray}
and they involve a trace over colour matrices, and a sum over Matsubara
frequencies. 

Again note that the extension of Furry's theorem to finite 
temperature does not hold 
at finite density: if we set $n \rightarrow -n-1$
we note that $ q_{0} \rightarrow \!\!\!\!\!\!\!\!\!/ \hspace{0.4mm} -q_{0} $ 
and as a result $T^{\mu \rho \nu} (\mu,T) \not= -T^{\nu \rho \mu} (\mu,T)$.
Of course, If we now let the chemical potential go to zero 
($\mu \rightarrow 0$) we note
that for the transformation  $n \rightarrow -n-1$  we obtain 
$q_{0} \rightarrow  -q_{0}$ and  thus 
$ T^{\mu \rho \nu} (0,T)  \rightarrow -T^{\nu \rho \mu} (0,T) $. The analysis for
fermion loops with larger number of vertices is essentially the same. 

\vspace{-0.5cm}
\subsection{Dilepton rates}

To calculate the contribution made by the diagram of
Fig. \ref{2vert}
to the dilepton spectrum emanating from a quark gluon plasma we 
calculate the imaginary part of the photon self-energy containing the above diagram
as an effective vertex (see \cite{maga2001} for details). 
We calculate in the limit of photon three momentum $\vec{p}=0$.
The imaginary part of the considered self energy contains various cuts. We concentrate
solely on the cut that represents the process of 
gluon-gluon to $\mbox{e}^{+}\mbox{e}^{-}$. This is a new process which to our
knowledge has not been discussed before. The other possible cuts represent
essentially finite density contributions to other known processes of dilepton 
production.
The differential production rate for pairs of massless leptons with total
energy $E$ and and total momentum $\vec{p}=0$ is given in terms of the discontinuity in the
photon self-energy \cite{gal91}. 
%
%
We use
the $q \bar{q} \to e^+ e^-$ Born process as a baseline. 
%

\begin{figure}[htbp]
  \resizebox{14pc}{!}{\includegraphics[height=.2\textheight]{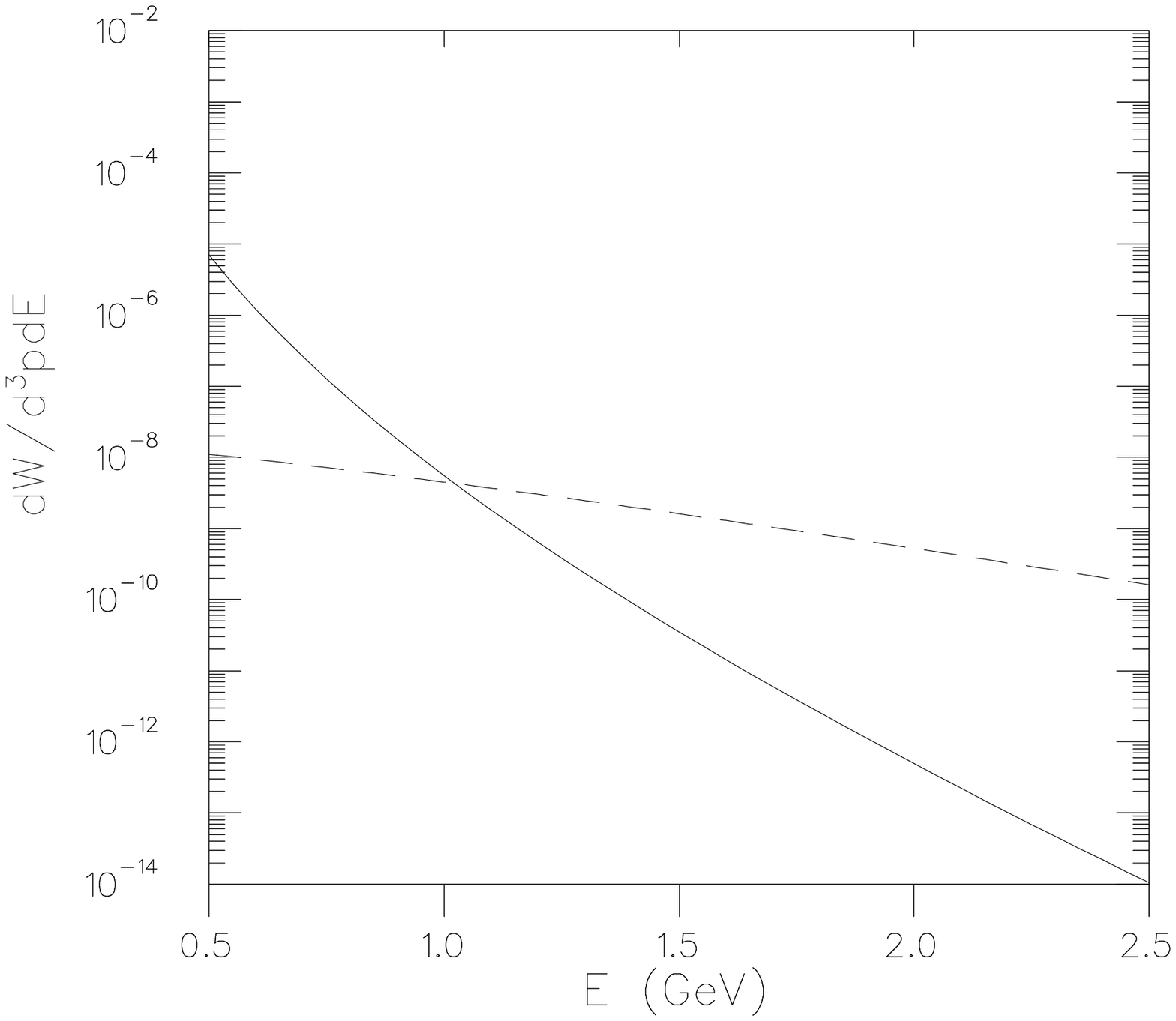}}
   \hspace{1cm}
   \resizebox{14pc}{!}{\includegraphics[height=.2\textheight]{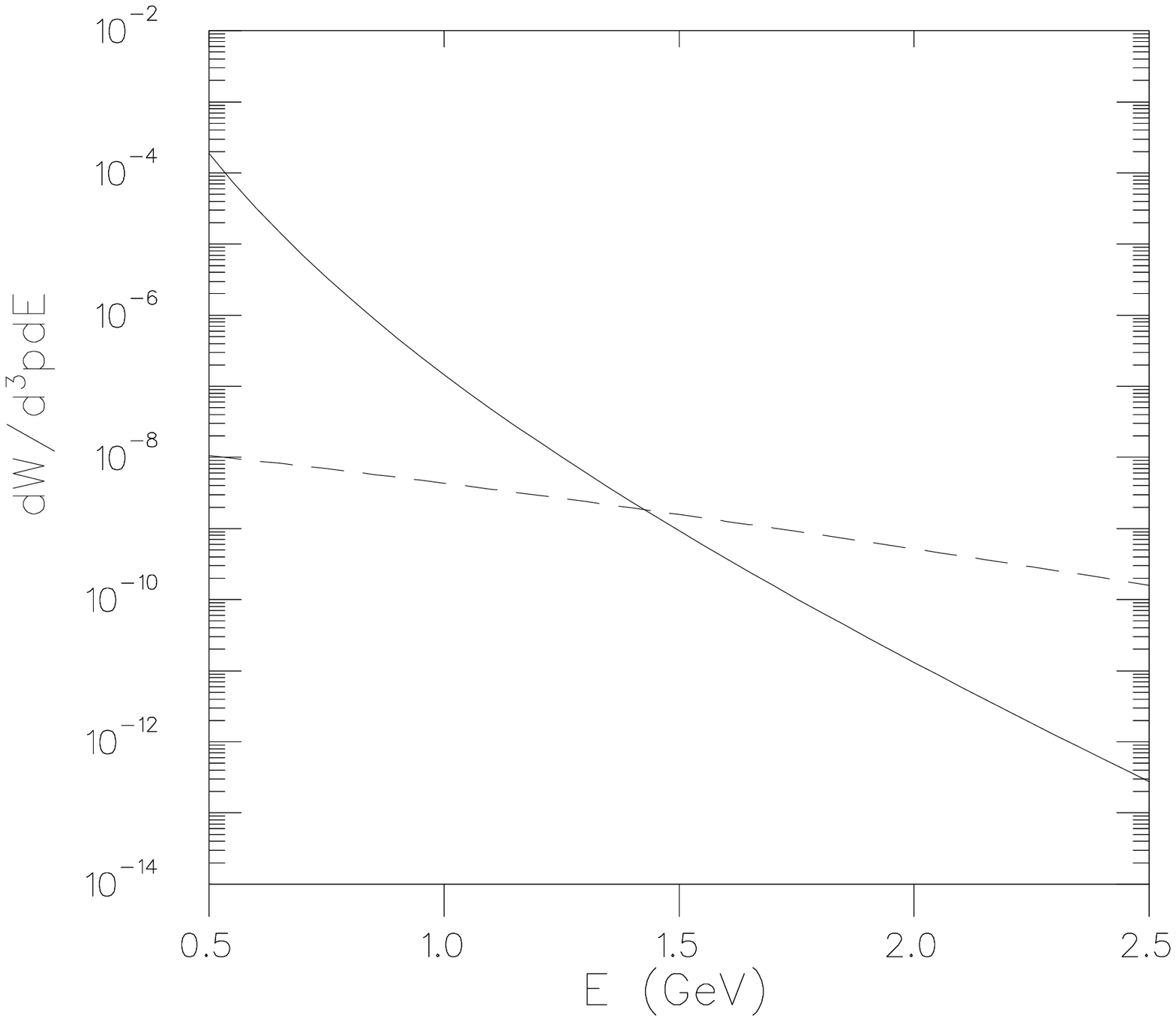}}
    \caption{The differential production rate of back to back dileptons
from two processes. Invariant mass runs from 0.5 GeV to 2.5 GeV.
The dashed line represents the contribution from the process 
$q\bar{q}\rightarrow e^{+}e^{-}$. The solid line corresponds to 
the process $gg\rightarrow e^{+}e^{-}$. Temperature is 400 MeV. 
Quark chemical potential is 0.1T. The second figure is the same as the first but with
 $\mu=0.5T$  }
    \label{mu}
\end{figure}

The initial temperatures of the plasma formed at RHIC and LHC 
have been estimated to lie in the range from 300-800 MeV \cite{wan96,rap00}.
 For this exploratory calculation 
we use a conservative estimate of $T=400\mbox{ MeV}$. To evaluate the effect
 of a finite chemical potential we perform the calculation  
with two extreme values of
chemical potential $\mu=0.1T$ (1st plot in Fig. \ref{mu}) and 
$\mu=0.5T$(2nd plot Fig. \ref{mu})
\cite{gei93}.
The calculation, is performed for three massless flavours of quarks. 
 The differential rate for the production of dileptons with an invariant mass from 
 0.5 to $2.5\mbox{ GeV}$ is presented. On purpose, we avoid regions where the gluons
 become very soft.
  In the plots, the dashed line is the rate from tree level
 $q\bar{q}$; the solid line is that from the process 
 $gg\rightarrow e^{+}e^{-}$.  We note that in both cases the gluon-gluon 
process dominates at low energy and dies out at higher energy leaving 
the $q\bar{q}$ process dominant. Importantly, the
crossing point lies near the continuum between the $\phi$ and the
$J/\psi$, raising an optimistic prospect for the successful detection of
the new contribution.

\vspace*{-0.5cm}
\section{Summary}

We have considered the lepton pair signature of symmetry breaking effects
in two different energy regimes. Those effects are present in finite
density media that break charge conjugation invariance. They 
vanish identically in vacuum,  are therefore genuine in-medium
manifestations, and are a testimony to the richness of the
many-body problem. Theoretical refinements are needed before truly
quantitative predictions can be made. In the lower energy  sector, 
many-body theory has to be introduced  in order to consistently extend 
the current vector meson spectral densities to include meson mixing
effects \cite{tdg}. At higher energies, we have begun a study of the effects of
Hard Thermal Loop resummation \cite{pis2,maj012}. Finally, our rates
will be inserted in appropriate dynamical models.  As those works are in
progress, it is stimulating to note that the signals discussed in this
work are all accessible to the new generation of lepton pair
measurements.

\vspace*{-.5cm}
\begin{theacknowledgments}

This work was supported in part by the Natural Sciences and Engineering
Research Council of Canada, and in part by the Fonds FCAR of the Quebec
government. 

\end{theacknowledgments}


\vspace*{-.5cm}
\begin{thebibliography}{99}
\bibitem{sak}J. J. Sakurai, {\it Currents and Mesons}, University of
Chicago Press, Chicago, 1969.

\bibitem{gal91} C. Gale and J. I. Kapusta, {\it Nucl. Phys.}, {\bf B357}, 
65 (1991).

\bibitem{rawa}R. Rapp and J. Wambach, {\it Adv. Nucl. Phys.}, {\bf 25}, 1
(2000), and references therein.

\bibitem{ceres}See H. Appelshaeuser, Proceedings of Quark Matter 2001,
{\it Nucl. Phys. A}, in press, and references therein.

\bibitem{gaqm} See, for example, Charles Gale,
Proceedings of Quark Matter 2001,
{\it Nucl. Phys. A}, in press, and references therein.
 
 
\bibitem{chin77} S. A. Chin, {\it Ann. Phys.}, {\bf 108}, 301, (1977)

 \bibitem{wolf98} G. Wolf, B. Friman, and M. Soyeur, {\it Nucl. Phys.},  
{\bf A640}, 129, (1998).

\bibitem{saito98} K. Saito, K. Tsushima, A. W. Thomas, and A. G. Williams,
{\it Phys. Lett.}, {\bf B433}, 243 (1998).

\bibitem {teodorescu00} O. Teodorescu, A.K. Dutt-Mazumder, and C. Gale,
{\it Phys.Rev. C},  {\bf 61}, 051901 (2000); {\it Phys. Rev. C}, {\bf
63}, 034903 (2001).

\bibitem{sewal} B. D. Serot and J. D. Walecka, {\it Adv. Nucl. 
Phys.}, {\bf 16}, 1 (1986).

\bibitem{maga2001}A. Majumder and C. Gale, {\it Phys. Rev. D}, {\bf 63}, 
114008 (2001), and erratum in press. 

\bibitem{hades}J. Friese {\it et al.}, {\it Prog. Part. Nucl. Phys.}, {\bf
42}, 235 (1999).

\bibitem{fur37} W. H. Furry, {\it Phys. Rev.}, {\bf 51}, 125 (1937).



\bibitem{wel92}H. A. Weldon, {\it Phys. Lett. B}, {\bf 274}, 133 (1992). 

 
\bibitem{wan96} X. N. Wang, {\it Phys. Rept.},  {\bf 280}, 287 (1997).
 
\bibitem{rap00} R. Rapp, hep-ph/0010101.

\bibitem{gei93} K. Geiger and J. I. Kapusta, {\it Phys. Rev. D}, 
{\bf 47}, 4905 (1993); N.
George, Proceedings of Quark Matter 2001,
{\it Nucl. Phys. A}, in press. 


\bibitem{tdg}O. Teodorescu, A. K. Dutt-Mazumder, and C. Gale, in
preparation.

\bibitem{pis2} E. Braaten and R. D. Pisarski, {\it Nucl. Phys.}, 
{\bf B337}, 569 (1990). 


\bibitem{maj012} A. Majumder and C. Gale, {\it Proceedings of 
MRST 2001 }, to appear, and in preparation.


\end{thebibliography}
\end{document}